\documentclass[12pt]{article}
\usepackage[dvips]{graphicx}
\usepackage{cite}
\usepackage{epsfig}
\usepackage{latexsym,amsmath,amsfonts,amssymb}
\usepackage{mathtools} 
\usepackage{slashed} 
\usepackage{blkarray}
\usepackage[all]{xy} 
\usepackage[makeroom]{cancel}
\usepackage[latin1]{inputenc}
\usepackage[american]{babel}
\usepackage[dvips]{graphicx}
\usepackage{bbm}
\usepackage{color}
\usepackage[unicode]{hyperref}
\def\im{Invent. Math.}

\def\a{\alpha}
\def\b{\beta}
\def\c{\gamma}
\def\d{\delta}
\def\f{\phi}               %      \varphi
\def\vf{\varphi}  \def\tvf{\tilde{\varphi}}
\def\vp{\varphi}
\def\g{\gamma}
\def\h{\eta}
\def\j{\psi}
\def\k{\kappa}                    % Also, \varkappa (see below)
\def\l{\lambda}
\def\m{\mu}
\def\n{\nu}
\def\o{\omega}  \def\w{\omega}

\def\q{\theta}  \def\th{\theta}                  %     \vartheta
\def\r{\rho}                                     %     \varrho
\def\s{\sigma}                                   %     \varsigma
\def\t{\tau}
\def\u{\upsilon}
\def\x{\xi}
\def\z{\zeta}
\def\pt{\tilde{\varphi}}
\def\tt{\tilde{\theta}}
\def\lab{\label}
\def\6{\partial}
\def\wg{\wedge}
\def\bpsi{\bar{\psi}}
\def\bt{\bar{\theta}}
\def\bvf{\bar{\varphi}}

\DeclareMathOperator{\AdS}{AdS}
\DeclareMathOperator{\tr}{tr}

\newcommand{\be}{\begin{equation}}
\newcommand{\ee}{\end{equation}}
\newcommand{\beq}{\begin{equation}}
\newcommand{\eeq}{\end{equation}}
\newcommand{\bea}{\begin{eqnarray}}
\newcommand{\eea}{\end{eqnarray}}
\newcommand{\nn}{\nonumber \\}

\newcommand{\ba}{\begin{eqnarray}}
\newcommand{\ea}{\end{eqnarray}}

\newcommand{\beqs}{\begin{eqnarray}}
\newcommand{\eeqs}{\end{eqnarray}}
\newcommand{\bal}{\begin{aligned}}
\newcommand{\eal}{\end{aligned}}

\begin{document}
\baselineskip=15.5pt
\pagestyle{plain}
\setcounter{page}{1}
%--------+---------+---------+---------+---------+---------+---------+
%Body

% Ofer's definitions

\def\del{{\partial}}
\def\vev#1{\left\langle #1 \right\rangle}
\def\cn{{\cal N}}
\def\co{{\cal O}}
%\newfont{\Bbb}{msbm10 scaled 1200}     %instead of eusb10
%\newcommand{\mathbb}[1]{\mbox{\Bbb #1}}
\def\IC{{\mathbb C}}
\def\IR{{\mathbb R}}
\def\IZ{{\mathbb Z}}
\def\RP{{\bf RP}}
\def\CP{{\bf CP}}
\def\Poincare{{Poincar\'e }}
\def\tr{{\rm tr}}
\def\tp{{\tilde \Phi}}

\def\TL{\hfil$\displaystyle{##}$}
\def\TR{$\displaystyle{{}##}$\hfil}
\def\TC{\hfil$\displaystyle{##}$\hfil}
\def\TT{\hbox{##}}
\def\HLINE{\noalign{\vskip1\jot}\hline\noalign{\vskip1\jot}}
\def\seqalign#1#2{\vcenter{\openup1\jot
   \halign{\strut #1\cr #2 \cr}}}
\def\lbldef#1#2{\expandafter\gdef\csname #1\endcsname {#2}}
\def\eqn#1#2{\lbldef{#1}{(\ref{#1})}%
\begin{equation} #2 \label{#1} \end{equation}}
\def\eqalign#1{\vcenter{\openup1\jot
     \halign{\strut\span\TL & \span\TR\cr #1 \cr
    }}}
\def\eno#1{(\ref{#1})}
\def\href#1#2{#2}
\def\half{\frac{1}{2}}

%--------+---------+---------+---------+---------+---------+---------+
%Hirosi's macros:
\def\ads{{\it AdS}}
\def\adsp{{\it AdS}$_{p+2}$}
\def\cft{{\it CFT}}

\newcommand{\ber}{\begin{eqnarray}}
\newcommand{\eer}{\end{eqnarray}}

\newcommand{\beqar}{\begin{eqnarray}}
\newcommand{\cN}{{\cal N}}
\newcommand{\cO}{{\cal O}}
\newcommand{\cA}{{\cal A}}
\newcommand{\cT}{{\cal T}}
\newcommand{\cF}{{\cal F}}
\newcommand{\cC}{{\cal C}}
\newcommand{\cR}{{\cal R}}
\newcommand{\cW}{{\cal W}}
\newcommand{\eeqar}{\end{eqnarray}}
\newcommand{\tht}{\thteta}
\newcommand{\lm}{\lambda}\newcommand{\Lm}{\Lambda}
\newcommand{\R}{\mathbb{R}}

%--------+---------+---------+---------+---------+---------+---------+

\newcommand{\nonu}{\nonumber}
\newcommand{\oh}{\displaystyle{\frac{1}{2}}}
\newcommand{\dsl}
   {\kern.06em\hbox{\raise.15ex\hbox{$/$}\kern-.56em\hbox{$\partial$}}}
\newcommand{\id}{i\!\!\not\!\partial}
\newcommand{\as}{\not\!\! A}
\newcommand{\ps}{\not\! p}
\newcommand{\ks}{\not\! k}
\newcommand{\D}{{\cal{D}}}
\newcommand{\dv}{d^2x}
\newcommand{\Z}{{\cal Z}}
\newcommand{\N}{{\cal N}}
\newcommand{\Dsl}{\not\!\! D}
\newcommand{\Bsl}{\not\!\! B}
\newcommand{\Psl}{\not\!\! P}
\newcommand{\eeqarr}{\end{eqnarray}}
\newcommand{\ZZ}{{\rm \kern 0.275em Z \kern -0.92em Z}\;}
%--------------------------------Alfonso's definitions%%%%%%%%%%%%%

% DEFINITIONS

\def\del{{\delta^{\hbox{\sevenrm B}}}} \def\ex{{\hbox{\rm e}}}
\def\azb{A_{\bar z}} \def\az{A_z} \def\bzb{B_{\bar z}} \def\bz{B_z}
\def\czb{C_{\bar z}} \def\cz{C_z} \def\dzb{D_{\bar z}} \def\dz{D_z}
\def\im{{\hbox{\rm Im}}} \def\mod{{\hbox{\rm mod}}} \def\tr{{\hbox{\rm Tr}}}
\def\ch{{\hbox{\rm ch}}} \def\imp{{\hbox{\sevenrm Im}}}
\def\trp{{\hbox{\sevenrm Tr}}} \def\vol{{\hbox{\rm vol}}}
\def\rl{\Lambda_{\hbox{\sevenrm R}}} \def\wl{\Lambda_{\hbox{\sevenrm W}}}
\def\fc{{\cal F}_{k+\cox}} \def\vev{vacuum expectation value}
\def\nodiv{\mid{\hbox{\hskip-7.8pt/}}}
\def\ie{{\em i.e.}}
\def\ie{\hbox{\it i.e.}}

\def\CC{{\mathchoice
{\rm C\mkern-8mu\vrule height1.45ex depth-.05ex
width.05em\mkern9mu\kern-.05em}
{\rm C\mkern-8mu\vrule height1.45ex depth-.05ex
width.05em\mkern9mu\kern-.05em}
{\rm C\mkern-8mu\vrule height1ex depth-.07ex
width.035em\mkern9mu\kern-.035em}
{\rm C\mkern-8mu\vrule height.65ex depth-.1ex
width.025em\mkern8mu\kern-.025em}}}

\def\RR{{\rm I\kern-1.6pt {\rm R}}}
\def\NN{{\rm I\!N}}
\def\ZZ{{\rm Z}\kern-3.8pt {\rm Z} \kern2pt}
\def\IB{\relax{\rm I\kern-.18em B}}
\def\ID{\relax{\rm I\kern-.18em D}}
\def\II{\relax{\rm I\kern-.18em I}}
\def\IP{\relax{\rm I\kern-.18em P}}
\newcommand{\CS}{{\scriptstyle {\rm CS}}}
\newcommand{\CSs}{{\scriptscriptstyle {\rm CS}}}
\newcommand{\rc}{\nonumber\\}
\newcommand{\bear}{\begin{eqnarray}}
\newcommand{\eear}{\end{eqnarray}}

\newcommand{\LL}{{\cal L}}

\def\mani{{\cal M}}
\def\calo{{\cal O}}
\def\calb{{\cal B}}
\def\calw{{\cal W}}
\def\calz{{\cal Z}}
\def\cald{{\cal D}}
\def\calc{{\cal C}}
\def\to{\rightarrow}
\def\ele{{\hbox{\sevenrm L}}}
\def\ere{{\hbox{\sevenrm R}}}
\def\zb{{\bar z}}
\def\wb{{\bar w}}
\def\nodiv{\mid{\hbox{\hskip-7.8pt/}}}
\def\menos{\hbox{\hskip-2.9pt}}
\def\dr{\dot R_}
\def\drr{\dot r_}
\def\ds{\dot s_}
\def\da{\dot A_}
\def\dga{\dot \gamma_}
\def\ga{\gamma_}
\def\dal{\dot\alpha_}
\def\al{\alpha_}
\def\cl{{closed}}
\def\cls{{closing}}
\def\vev{vacuum expectation value}
\def\tr{{\rm Tr}}
\def\to{\rightarrow}
\def\too{\longrightarrow}

% Umut likes:

\def\a{\alpha}
\def\b{\beta}
\def\c{\gamma}
\def\d{\delta}
\def\e{\epsilon}           % Also, \varepsilon
\def\F{\Phi}
\def\f{\phi}               %      \varphi
\def\vf{\varphi}  \def\tvf{\tilde{\varphi}}
\def\vp{\varphi}
\def\g{\gamma}
\def\h{\eta}
\def\j{\psi}
\def\k{\kappa}                    % Also, \varkappa (see below)
\def\l{\lambda}
\def\m{\mu}
\def\n{\nu}
\def\o{\omega}  \def\w{\omega}
\def\q{\theta}  \def\th{\theta}                  %     \vartheta
\def\r{\rho}                                     %     \varrho
\def\s{\sigma}                                   %     \varsigma
\def\t{\tau}
\def\u{\upsilon}
\def\x{\xi}
\def\X{\Xi}
\def\z{\zeta}
\def\pt{\tilde{\varphi}}
\def\tt{\tilde{\theta}}
\def\lab{\label}
\def\6{\partial}
\def\wg{\wedge}
\def\atanh{{\rm arctanh}}
\def\bpsi{\bar{\psi}}
\def\bt{\bar{\theta}}
\def\bvf{\bar{\varphi}}

%
% FONTS

%\newfont{\headfont}{cmbx10 scaled 1440}
\newfont{\namefont}{cmr10}
%\newfont{\initialfont}{cmr10 scaled 1200}
\newfont{\addfont}{cmti7 scaled 1440}
\newfont{\boldmathfont}{cmbx10}
%\newfont{\figfont}{cmr7 scaled 1200}
\newfont{\headfontb}{cmbx10 scaled 1728}
%%%%%%%%%%%%%%%%%%%%%%%%%%%%%%%%%%%%%%%%%%%%%%%%%%%%%%%%%%%%%
%%%%%%%%%%%%%%%
%%%%%%%%%%%%%%%Stefano and Francesco fonts%%%%%%%%%%%%%%%%%%%%%%%%%%%%%%%%
%%%%%%%%%%%%%%%%%%%%%%%%%%%%%%%%%%%%%%%%%%%%%%%%%%%%%%%%%%%%%
%%%%%%%%%%%%%%%
\newcommand{\re}{\,\mathbb{R}\mbox{e}\,}
\newcommand{\hyph}[1]{$#1$\nobreakdash-\hspace{0pt}}
\providecommand{\abs}[1]{\lvert#1\rvert}
\newcommand{\Nugual}[1]{$\mathcal{N}= #1 $}
\newcommand{\sub}[2]{#1_\text{#2}}
\newcommand{\partfrac}[2]{\frac{\partial #1}{\partial #2}}
\newcommand{\bsp}[1]{\begin{equation} \begin{split} #1 \end{split} \end{equation}}
\newcommand{\calF}{\mathcal{F}}
\newcommand{\calO}{\mathcal{O}}
\newcommand{\calM}{\mathcal{M}}
\newcommand{\calV}{\mathcal{V}}
\newcommand{\bbZ}{\mathbb{Z}}
\newcommand{\bbC}{\mathbb{C}}
\newcommand{\cK}{{\cal K}}
\newcommand{\dd}{\textrm{d}}
\newcommand{\DD}{\textrm{D}}

%%%%%%% Definitions for the current file
%\newcommand{\td}{\mathrm{d}}
%\newcommand{\vol}[1]{\textrm{Vol(}#1\textrm{)}}
\newcommand{\Thq}{\Theta\left(\r-\r_q\right)}
\newcommand{\Dq}{\d\left(\r-\r_q\right)}
\newcommand{\kten}{\kappa^2_{\left(10\right)}}
\newcommand{\pbi}[1]{\imath^*\left(#1\right)}
\newcommand{\ho}{\hat{\omega}}
\newcommand{\tth}{\tilde{\th}}
\newcommand{\tf}{\tilde{\f}}
\newcommand{\tj}{\tilde{\j}}
\newcommand{\tw}{\tilde{\omega}}
\newcommand{\tz}{\tilde{z}}
\newcommand{\prj}[2]{(\partial_r{#1})(\partial_{\j}{#2})-(\partial_r{#2})(\partial_{\j}{#1})}
\def\atanh{{\rm arctanh}}
\def\sech{{\rm sech}}
\def\csch{{\rm csch}}
\allowdisplaybreaks[1]

\def\red{\textcolor[rgb]{0.98,0.00,0.00}}

\numberwithin{equation}{section}

\newcommand{\Tr}{\mbox{Tr}}    % trace over gauge indices

%%%%%%%%%%%%%%%%%%%%%%%%%%%%%%%%%%%%%%%%%%%%%%%%%%%%%%%%%%%%%
%%%%%%%%%%%%%%%
%%%%%%%%%%%%%%%%%%%%%%%%%%%%%%%%%%%%%%%%%%%%%%%%%%%%%%%%%%%%%
%%%%%%%%%%%%%%%

%
\renewcommand{\theequation}{{\rm\thesection.\arabic{equation}}}
\begin{titlepage}

\vfill
\begin{flushright}
APCTP Pre2018-004 
\end{flushright}

\vfill

\begin{center}
   \baselineskip=16pt
   {\Large \bf Embedding the modified CYBE in Supergravity }
   \vskip 2cm
   Thiago Araujo$^a$, Eoin \'O Colg\'ain$^a$ \& Hossein Yavartanoo$^b$
       \vskip .6cm
             \begin{small}
               \textit{$^a$ Asia Pacific Center for Theoretical Physics, Postech, Pohang 37673, Korea}
               
               \vspace{3mm} 
               
               \textit{$^b$ State Key Laboratory of Theoretical Physics, Institute of Theoretical Physics, \\ Chinese Academy of Sciences, Beijing 100190, China}
  
             \end{small}
\end{center}

\vfill \begin{center} \textbf{Abstract}\end{center} \begin{quote}
It has recently been demonstrated that given a generic solution, the Classical Yang-Baxter Equation (CYBE) emerges from supergravity via the open-closed string map, thus providing tangible evidence for the conjectured equivalence between supergravity equations of motion and the homogeneous CYBE. To date, study of this equivalence has largely been confined to the NS sector. In this work, we make two extensions. First, we revisit the transformation of the RR sector and clarify its precise role in the emergence of the CYBE. Secondly, we identify direct products of coset geometries as the only setting where the transformation permits embeddings of the modified CYBE. We illustrate our solution generating technique with deformations of $AdS_3 \times S^3 \times M_4$, where $M_4 = T^4$ ($K3$) and $S^3 \times S^1$, and explicitly construct one and two-parameter integrable q-deformations that are solutions to generalised supergravity. 
\end{quote} \vfill

\end{titlepage}

\setcounter{footnote}{0}
\renewcommand{\theequation}{{\rm\thesection.\arabic{equation}}}
\section{Introduction}
Over recent years, the Yang-Baxter $\sigma$-model \cite{Klimcik:2002zj, Delduc:2013qra, Kawaguchi:2014qwa} has emerged as a systematic way to construct integrable deformations of maximally symmetric AdS/CFT geometries. In principle this extends the scope of integrability techniques in holography to more realistic settings. Central to this approach is an $r$-matrix solution to the Classical Yang-Baxter Equation (CYBE). The Yang-Baxter $\sigma$-model was initially formulated in terms of $r$-matrix solutions to the \textit{modified} CYBE, before it was later understood that there was a richer class of deformations based on $r$-matrix solutions to the \textit{homogeneous} CYBE \cite{Kawaguchi:2014fca, Matsumoto:2014nra, Matsumoto:2014gwa}. 

Since there are fewer solutions to the modified CYBE, it is perfectly understandable that the corresponding supergravity solutions are rarer \footnote{See \cite{Delduc:2014kha, Hoare:2014pna, Hoare:2016ibq} for a discussion of inequivalent $AdS_n \times S^n$ deformations.}. However, this appears to be a disproportionate rareness, since given an $r$-matrix solution to the modified CYBE, there is no guarantee a corresponding embedding in supergravity exists. This ultimately can be traced to the fact that such solutions are less natural from the supergravity perspective, as we will explain in due course.  

We recall from a series of recent papers \cite{Araujo:2017jkb, Araujo:2017jap, Araujo:2017enj} that Yang-Baxter deformations for $r$-matrices based on \textit{both} the homogeneous and modified CYBE are described by an open-closed string map \cite{Seiberg:1999vs}, where the deformation is specified by a bivector $\Theta$ that is an antisymmetric product of Killing vectors, or simply ``bi-Killing". The joy of this map is it reduces the deformation to a single matrix inversion in the $\sigma$-model target space. Moreover, the map can be built into a powerful solution generating technique \cite{Bakhmatov:2017joy} for generalised supergravity \cite{Arutyunov:2015mqj, Wulff:2016tju}. This approach hinges on the assumption, explicitly checked case by case in \cite{Bakhmatov:2017joy, Bakhmatov:2018apn}, that the equations of generalised supergravity reduce to the CYBE, and once the \textit{homogeneous} CYBE holds, so too do the equations of motion. In the process, $\Theta$ is identified with an $r$-matrix solution \footnote{See \cite{vanTongeren:2015uha, vanTongeren:2016eeb} for earlier comments linking $\Theta$ with $r$-matrices and \cite{Sakamoto:2017cpu, Fernandez-Melgarejo:2017oyu, Sakamoto:2018krs} where the open-closed string map is recast as a $\beta$-transformation in the context of $O(d,d)$ transformations.}. 

A step towards a proof of the equivalence between the equations of motion of generalised supergravity and the CYBE for deformations appeared in \cite{Bakhmatov:2018apn}. In particular, restricting to the NS sector, and geometries not supported by the NSNS two-form, or $B$-field, it was shown perturbatively that the equations of motion are equivalent to the CYBE once $\Theta$ is bi-Killing. More precisely, it has been checked to a certain order that the equations of motion hold once the homogeneous CYBE is satisfied. This leaves valid questions regarding generalisations to i) supergravity solutions supported by RR field strengths and ii) examples based on the modified CYBE, both of which fall outside the perturbative analysis  \cite{Bakhmatov:2018apn}. In this work, we continue our study in this direction and provide further examples of the solution generating technique \cite{Bakhmatov:2017joy} in a bid to address both these questions. We emphasise that our method to construct the solutions, versus T-duality on non-isometric directions \cite{Hoare:2015wia} (see also \cite{Orlando:2016qqu}), is more direct as we work in generalised supergravity from the outset. 

To address the first issue, in section \ref{sec:method} we outline the role of the RR sector in the solution generating technique of \cite{Bakhmatov:2017joy} by making use of an illustrative example. We note that the RR sector simply supports geometries and the CYBE emerges exclusively from the NS sector. This should come as little surprise, since our prescription for deforming the RR sector \cite{Bakhmatov:2017joy} (see section \ref{sec:method}) does not involve derivatives of $\Theta$ and is thus insensitive to the actual make-up of the bivector defining the deformation. Put differently, the RR sector sees little difference between a TsT transformation \cite{Lunin:2005jy}, which trivially satisfies the homogeneous CYBE, and more involved Yang-Baxter deformations. 

As for our second concern, namely geometries based on the modified CYBE, it is instructive to recall the $\kappa$-deformation of the Poincar\'e algebra $\mathfrak{g} = \mathfrak{iso}(1,d)$, which is specified by the $r$-matrix, 
\be
\label{r-matrix}
r = a^{\mu} M_{\mu \nu} \wedge P^{\nu}, 
\ee
where $a$ is a constant vector. When $a$ is light-like (null), it is well documented, e. g.  \cite{Borowiec:2013lca, Borowiec:2015wua},  that the $r$-matrix is a solution to the homogeneous CYBE, otherwise we get a solution to the modified CYBE. As we show in the appendix, starting from flat spacetime, the light-like $\kappa$-deformation leads to a ``trivial" solution \cite{Wulff:2018aku} of generalised supergravity \footnote{Since we are deforming flat spacetime, this provides arguably the simplest example in this class.}. In contrast, we find using our method it is not possible to generate a deformation of flat spacetime that admits $\kappa$-deformations where $a$ is not null \footnote{As we explain later, our prescription for generating solutions based on the modified CYBE rests upon displacing the NS and RR sectors. Without an RR sector, this prescription does not work. However, it is possible to find deformations of flat spacetime by contracting $AdS_5 \times S^5$ deformations \cite{Pachol:2015mfa}.}. 

This above example is instructive as it illustrates how it is relatively easy to generate supergravity solutions based on $r$-matrix solutions to the homogeneous CYBE, whereas this is not immediate for their modified CYBE counterparts. This comes down to the fact that homogeneous Yang-Baxter deformations are more natural from the perspective of supergravity, since as we explained, once the bi-Killing deformation parameter satisfies the homogeneous CYBE, the equations of motion hold. This naturalness is also evident in other results, in particular the fact that homogeneous Yang-Baxter deformations may be understood as non-Abelian T-duality transformations \cite{Hoare:2016wsk, Borsato:2016pas, Hoare:2016wca, Borsato:2017qsx}. It is telling that there is no interpretation of modified deformations as a T-duality transformation.  

In this work, to better understand when we may expect a supergravity solution based on the modified CYBE, we recycle the perturbative results quoted in \cite{Bakhmatov:2018apn}. From the dilaton equation, we see that one must consider multiple deformations whose contributions cancel amongst themselves, so that the equation is satisfied. This gives rise to a feeling of ``awkwardness", since we are in essence fitting a square plug in a round hole. Furthermore, allowing for the constant shift in the dilaton, we note from the Einstein equation that this shift must back-react on the entire geometry in cases where the stress-energy tensor does not vanish, i. e. non-flat directions \footnote{In string theory, a constant shift in the dilaton does not change the world-sheet action as the two-dimensional Ricci scalar is a topological number. This constant shift changes the effective string coupling constant and this is a modulus of the theory. It is also worth stressing that in the absence of an RR sector, the dilaton shift does not back-react.}. This means that for a curved spacetime, we have to deform all of the directions with no exceptions. Since our deformation makes use of Killing vectors, this can only be done where there is an isometry associated to each direction. This ultimately precludes warped-product spacetimes, e. g. Dp-brane geometries, and restricts us to deformations based on direct-products of coset geometries. This in turn implies that we should be able to deform geometries such as $AdS_5 \times T^{1,1}$, even though the motion of strings in this background is not classically integrable \cite{Basu:2011di}.  We flesh out these arguments in section \ref{sec:method}, where we also review the solution generating technique of \cite{Bakhmatov:2017joy} and elucidate the role of the RR sector. The above observation provides a supergravity insight into so-called $\eta$-deformations \cite{Delduc:2013qra, Delduc:2014kha} and explains why they should be restricted to cosets. 

As we will demonstrate, a key point of this work is that the method originally outlined in \cite{Bakhmatov:2017joy}, which has not been extensively tested in the literature, works and we can have greater confidence in its use. In particular, through Page forms and our descent procedure, one can simply write down solutions. Admittedly, the RR sectors can be involved, so here we focus on the geometries  $AdS_3 \times S^3 \times M_4$ deformations, where $M_4 = T^4$ or $K3$, and $M_4 = S^3 \times S^1$. It is worth noting that a two-parameter q-deformation of this geometry, considered earlier in \cite{Hoare:2014oua, Lunin:2014tsa}, can be extended to a full solution including the RR sector using our methods (section \ref{sec:ads3s3}) and that this deformation can be easily extended to a one or two-parameter q-deformation where the geometry encapsulates the exceptional Lie superalgebra $D(2, 1 ; \alpha)$ . In the case of the latter, we confirm that the deformation respects a well-known constraint on the radii of the symmetric spaces. The deformations are expected to preserve integrability, but being solutions to generalised supergravity, and not usual supergravity, the AdS/CFT interpretation of the deformed solutions is not clear. If one can be found, it is expected to be related to a noncommutative deformation of Yang-Mills \cite{vanTongeren:2015uha, vanTongeren:2016eeb, Araujo:2017jkb, Araujo:2017jap, Araujo:2017enj}.

\section{Methodology}
\label{sec:method}
We begin by reviewing the method outlined in \cite{Bakhmatov:2017joy}, which is valid for supergravity solutions not supported by an NSNS two-form \footnote{At the heart of our approach is the open-closed string map of Seiberg \& Witten \cite{Seiberg:1999vs} and in open string parameters there is no $B$-field. The extension to include the $B$-field, even in the presence of singular matrices $g+B$, is an open problem. See \cite{Sakamoto:2018krs} for progress in this direction. }.  Central to our prescription is the open-closed string map \cite{Seiberg:1999vs} 
\be
\label{open_closed}
(G^{-1} + \Theta )^{-1} = g + B, 
\ee
where $G$ denotes the original metric, $\Theta$ is a deformation parameter, essentially an antisymmetric bivector,  $g$ and $B$ correspond to the deformed metric and $B$-field. We assume that the original metric admits an isometry group, which allows us to define $\Theta$ as an antisymmetric product of Killing vectors $K^{\mu}_i$
\be
\Theta^{\mu \nu} = r^{ij} K^{\mu}_i K^{\nu}_j, 
\ee
where $r^{ij}$ is a skew-symmetric matrix, $r^{ij} = - r^{ji}$, with constant entries. One should note that the following vector,  
\be
\label{I_theta}
I^{\mu} = \nabla_{\nu} \Theta^{\nu \mu} \equiv \frac{1}{\sqrt{G}} \partial_{\nu} \left( \sqrt{G} \Theta^{\nu \mu} \right),
\ee
is Killing by construction and for non-zero $I$ it has been proved \cite{Bakhmatov:2018apn} that the deformed geometry corresponds to a solution to generalised supergravity \cite{Arutyunov:2015mqj}, where $I$ is the Killing vector modification of usual supergravity \cite{Arutyunov:2015mqj, Wulff:2016tju}. To complete the deformation in the NS sector, the scalar dilaton of the deformed solution $\phi$ is related to the original dilaton $\Phi$ through a well-known T-duality invariant, 
\be
\label{tduality_inv}
e^{- 2 \Phi} \sqrt{G} = e^{-2 \phi} \sqrt{g}. 
\ee
For deformations involving just the NS sector of supergravity, it has been shown for generic spacetime metrics that the CYBE emerges at second order from the equations of motion \cite{Bakhmatov:2018apn}. Given an explicit solution, one can go further and show that the equations of motion reduce to the CYBE and the coefficients of $\Theta$ are constrained so that it is an $r$-matrix solution to the (homogeneous) CYBE \footnote{See \cite{Bakhmatov:2017joy} for deformations of the Schwarzschild metric.}. It should be noted that in contrast to the Yang-Baxter $\sigma$-model, which typically involves deformations of coset geometries and is purely algebraic, our method reduces to the single matrix inversion (\ref{open_closed}). It can be checked that the methods are equivalent \cite{Araujo:2017jkb, Araujo:2017jap, Araujo:2017enj}.

With a prescription for deforming the NS sector in hand, it is pretty immediate to extend this to the RR sector. The traditional approach is to identify how the transformation acts on spinors and use this knowledge to reconstruct the transformation of the RR sector \cite{Hassan:1999bv, Kelekci:2014ima, Borsato:2016ose}. In \cite{Bakhmatov:2017joy} an alternative way was suggested. This method is rooted in the philosophy that all information of the deformation must follow from a knowledge of $\Theta$. In contrast to the usual approach, we arrive at the deformed RR sector by descent simply by contracting $\Theta$ into Page forms. This technique is based on the observation made originally in \cite{Araujo:2017enj} that the equations of motion of the RR sector, when re-expressed in terms of Page forms (\ref{Page}), take the simple form (\ref{dQQ}). 

Observe that when $I=0$ we recover the equations of motion of usual type II supergravity where the Page forms are closed and can be quantised as a result. Our prescription now demands that for each non-zero Page form $\tilde{Q}$ from the original supergravity solution, which is guaranteed to be closed, we define induced Page forms $Q$ associated to this form by descent \footnote{Given a $p$-form $A$ and $q$-form $B$ with $p \leq q$, we define the contraction $(A \lrcorner B)_{\mu_{p+1} \dots \mu_q} = \frac{1}{p!} A^{\mu_1 \dots \mu_p} B_{ \mu_1 \dots \mu_p \mu_{p+1} \dots \mu_q}$.}
\be
Q_{2(n-p)+1} = \frac{(-1)^p}{p!} \Theta^p \lrcorner \tilde{Q}_{2 n+1}.  
\ee
Summing up all the induced and original Page forms of a given degree, one finds a final expression for the final deformed Page form. It is easy to convince oneself using (\ref{I_theta}) and the closure of $\tilde{Q}$ that the RR equations of motion (\ref{dQQ}) are satisfied by construction. 

Therefore, the task of confirming a solution exists reduces to checking the Einstein (\ref{general1}), $B$-field (\ref{general2}) and dilaton (\ref{general3}) equations of motion, essentially the same equations one needs to check without an RR sector. In the absence of an RR sector, it has been shown perturbatively that a necessary condition for a deformed solution to exist is that $\Theta$ is an $r$-matrix solution to the homogeneous CYBE  \cite{Bakhmatov:2018apn}. Working with explicit solutions, one can show case by case that this condition is in fact sufficient. It should be clear that our prescription for the RR sector, in contrast to the NS sector, does not involve derivatives of $\Theta$ and for this reason the RR sector is ambivalent to the precise form of $\Theta$: it is ultimately determined exclusively by the NS sector. For this reason, the knowledge of the CYBE is encoded only in the NS sector and it is merely the role of the RR sector to support the geometry. This observation should come as no surprise, since if we had adopted the traditional approach and identified the transformation on spinors, we could use this knowledge to extract the transformed RR sector. 

As the above treatment may have been a tad abstract, let us review a known deformation of the $AdS_2 \times S^2 \times T^6$ geometry, which was identified using the above prescription \cite{Bakhmatov:2017joy}. We begin with the original intersecting D3-brane geometry, and in particular its near-horizon: 
\bea
\dd s^2 &=& \frac{ (- \dd t^2  + \dd z^2)}{z^2} + \dd \theta^2 + \sin^2 \theta \dd \phi^2 + \dd s^2 (T^6), \nn
F_{5} &=& (1 + *_{10}) \frac{1}{\sqrt{2} z^2} \dd t \wedge \dd z \wedge ( \omega_r - \omega_i), 
\eea
where $\omega_r, \omega_i$ are related to a complex $(3,0)$-form on the torus, $\omega_r - i \omega_i = \Omega_{(3,0)}$. With a view to teasing apart the role of the NS and RR sectors, we will revisit and dissect the equations of motion to see how they are satisfied. For clarity we focus only on the $AdS_2$ deformation. 

The $AdS_2$ Killing vectors are,  
\be
\label{KV}
K_1 = - t \partial_t - z \partial_z, \quad K_2  = - \partial_t, \quad K_3 = - (t^2 + z^2) \partial_t - 2 t z \partial_{z}, 
\ee
so the most general deformation parameter we can construct takes the form, 
\be
\Theta = \alpha K_1 \wedge K_2 + \beta K_2 \wedge K_3 + \gamma K_3 \wedge K_1 = \left(- \alpha z + \beta 2 t z + \gamma z ( - t^2 + z^2) \right) \partial_{t} \wedge \partial_{z}. 
\ee

Using the above prescription (\ref{open_closed}), (\ref{I_theta}) and (\ref{tduality_inv}), the deformed NS sector is easy to express as
\bea
\dd s^2 &=& \frac{z^2}{(z^4- \zeta^2)} (-\dd t^2 + \dd z^2), \quad B =  \frac{\zeta}{(z^4-\zeta^2)} \dd t \wedge \dd z, \nn
\phi &=& - \frac{1}{2} \log \left[ \frac{(z^4-\zeta^2)}{z^4} \right], \quad I =  \alpha K_2 - 2 \beta K_1 + \gamma K_3, 
\eea
where everything depends on a single function, 
\be
\zeta (t, z) \equiv - \alpha z + \beta 2 t z + \gamma z ( - t^2 + z^2). 
\ee
The rest of the geometry is undeformed and we omit it. Since we are working with a 2D geometry, $B$ is closed and as a result its field strength is zero, $H = 0$. This simplifies the equations of motion. Despite $H$ being zero, $B$ makes a contribution to the equations of motion through $X$ (\ref{X}) and in contrast to earlier work \cite{Arutyunov:2015mqj}, we cannot gauge it away if our map is to work. 

Contracting in $\Theta$, it is easy to identify the deformed Page forms by descent from the original Page five-form, before unravelling the Page forms to identify the RR field strengths in terms of $\zeta$:
\bea
F_5 &=& ( 1+ *_{10}) \frac{z^2}{\sqrt{2} (z^4-\zeta^2)} \dd t \wedge \dd z \wedge (\omega_r - \omega_i), \quad F_3 = - \frac{\zeta}{\sqrt{2} z^2} ( \omega_r - \omega_i). 
\eea
One can quickly confirm that the original geometry is recovered when $\zeta = 0$. 

Before going further, let us pause to comment on the solution and its implications for generalised IIA supergravity. We can envisage performing three T-dualities on the tori directions so that we are left with a zero-form (scalar) ``field strength" in type IIA supergravity. These T-dualities will not affect the geometry, just change the nature of the fluxes and the chirality of the theory. From the perspective of usual IIA supergravity, this scalar is expected to be the constant Romans' mass, however here we recognise that it is no longer a constant. Thus, it appears to be a generic feature of generalised IIA supergravity that the zero-form is not closed and has no interpretation as a mass. For completeness, in the appendix based on the T-dual of the above geometry, we write down the equations of motion it must satisfy. 

We now redirect our attention to the equations of motion. Since the equations involving the RR field strengths (\ref{dQQ}) are satisfied from the outset, we are left to consider the Einstein, $B$-field and dilaton equations. 

Let us start with the Einstein equation and decompose it in terms of NS and RR sector contributions. Owing to the symmetry in the $(t, x)$-directions, it is enough to focus on the temporal direction. The NS sector contribution is  
\be
\label{Einstein_ads2}
R_{tt} + 2 \nabla_{t} X_{t}  = \frac{z^2 (1 - 4 \beta^2 + 4 \alpha \gamma ) (z^4 + \zeta^2)}{(z^4 - \zeta^2)^2}, 
\ee
while the RR sector appears through the stress-energy tensor,  
\be
T_{tt} = \frac{z^2 (z^4 + \zeta^2)}{(z^4 - \zeta^2)^2}. 
\ee
One immediately recognises that the expressions are the same once $\beta^2 = \alpha \gamma$, which is precisely the homogeneous CYBE for the algebra $\frak{g} = \frak{sl}(2)$ once $\Theta$ is identified with the corresponding $r$-matrix solution to the CYBE. This term is traceable to derivatives of $\Theta$, or $\zeta$ in this case, and for TsT transformations where $\Theta$ is a constant in the usual frame, we see that the Einstein equation would have been satisfied. It is clear from this example that the non-trivial information about the CYBE is not coming from the RR sector, but instead from the NS sector. This conforms with our expectation that the RR sector acts largely as a spectator and has no direct bearing on the emergence of the CYBE. 

Before studying the other equations and arriving at the same conclusion, it is of interest to compare the above CYBE contribution to the NS sector with the expression derived in \cite{Bakhmatov:2018apn}. Since \cite{Bakhmatov:2018apn} studied only the NS sector equations of motion, we cannot expect to recover all the deformation terms at second order. From (3.12) of \cite{Bakhmatov:2018apn}, we recall the contribution to the Einstein equation at second order in $\Theta$, 
\be
\label{Einstein_2nd}
E^{(2)}_{\mu \nu} =  \frac{1}{2}  ( {\nabla}_{\rho} K_{ i \mu}  K_{ j \nu} K^{\rho}_{k}  + {\nabla}_{\rho} K_{ i \nu}  K_{ j \mu} K^{\rho}_{k}) \left( c_{l_1 l_2}^{~~~i} r^{j l_1} r^{k l_2} + c_{l_1 l_2}^{~~~k} r^{i l_1} r^{j l_2} + c_{l_1 l_2}^{~~~j} r^{k l_1} r^{i l_2}\right). 
\ee
Using the fact that the CYBE term is antisymmetric in indices and the commutation relations for the Lie algebra, we can rewrite this term as
\bea
E^{(2)}_{\mu \nu} &=& \frac{1}{2} (K_{[k \nu} c_{ij]}^{~~l} K_{l \mu} + K_{[k \mu} c_{ij]}^{~~l} K_{l \nu} ) \left( c_{l_1 l_2}^{~~~i} r^{j l_1} r^{k l_2} + c_{l_1 l_2}^{~~~k} r^{i l_1} r^{j l_2} + c_{l_1 l_2}^{~~~j} r^{k l_1} r^{i l_2}\right). 
%&=&  \frac{1}{12}  \biggl(  K_{k \nu} c_{ij}^{~~l} K_{l \mu} + K_{j \nu} c_{ki}^{~~l } K_{l \mu} + K_{i \nu} c_{jk}^{~~l} K_{l \mu} +  K_{k \mu} c_{ij}^{~~l} K_{l \nu} + K_{j \mu} c_{ki}^{~~l } K_{l \nu} \nn 
%&+& K_{i \mu} c_{jk}^{~~l} K_{l \nu} \biggr) \left( c_{l_1 l_2}^{~~~i} r^{j l_1} r^{k l_2} + c_{l_1 l_2}^{~~~k} r^{i l_1} r^{j l_2} + c_{l_1 l_2}^{~~~j} r^{k l_1} r^{i l_2}\right)
\eea
Evaluating this for the $E_{tt}$ component using the Killing vectors (\ref{KV}), we obtain 
\be
E_{tt}^{(2)} = \frac{4}{z^2} ( - \beta^2 + \alpha \gamma), 
\ee
which precisely agrees with the CYBE term in (\ref{Einstein_ads2}). Thus, if the perturbative analysis of \cite{Bakhmatov:2018apn} was extended to the RR sector, we can expect the same term (\ref{Einstein_2nd}) to appear at second order. We will return to this point soon. 

Let us move onto the $B$-field equation to confirm this picture. Once again, we decompose the equation into an NS,  
\be
\dd X = - \frac{2 z^4 \zeta (1 -  4 \beta^2 + 4 \alpha \gamma )}{(z^4-\zeta^2)^2} \dd t \wedge \dd z, 
\ee
and RR contribution, 
\be
\frac{1}{2} e^{2 \phi} F_3 \lrcorner F_5 = - \frac{2 z^4 \zeta}{(z^4-\zeta^2)^2} \dd t \wedge \dd z.  
\ee 
We see again that the contribution from the CYBE comes through the NS sector and not through the RR sector. Finally, the dilaton equation reads, 
\be
-\beta^2 + \alpha \gamma = 0, 
\ee
which once again is nothing more than the homogeneous CYBE. Here, we remark that the dilaton equation is an equation involving only the NS sector with the RR sector dropping out. So based on this example, we arrive at the following conclusion: the RR sector is merely a by-stander. It involves no derivatives acting on $\Theta$ and has no influence on the permitted deformations, but simply supports the original and deformed geometry. 

With a view to introducing the main focus of this paper, namely solutions based on the modified CYBE, it is pretty obvious that there is a second way to solve the above equations, which is simply to displace the RR sector relative to the NS sector. Since the RR sector appears in the equations dressed with the dilaton \footnote{In fact once the RR sector is solved by descent, the contribution of the RR sector to the remaining equations is the same as usual supergravity.}, it makes no difference if one shifts the dilaton by a constant or rescales the RR sector, the result is the same. We will take the view that we can shift the dilaton by an additional constant, $\phi \rightarrow \phi + \phi_0$.  From the Einstein equation and the $B$-field equation of motion, we see that choosing
\be
e^{2 \phi_0} =  1 - 4 \beta^2 + 4 \alpha \gamma
\ee
ensures that both equations are satisfied without constraining $\alpha, \beta$ and $\gamma$. Of course this constant shift in the dilaton does not affect the dilaton equation and this necessitates we perform some gymnastics to find an additional contribution to cancel  the dilaton equation terms, which are no longer zero. The solution to this is to also deform the two-sphere as explained in \cite{Bakhmatov:2017joy}.  Note, the same conclusion is reached by studying the Einstein equation in the $(\theta, \phi)$-directions, as once the dilaton is shifted, it back-reacts on the geometry and the Einstein equations in these directions will no longer be satisfied. In summary, it is clear that we have to also deform the two-sphere. 

This brings us back to the perturbative results of \cite{Bakhmatov:2018apn}, which are valid for generic spacetimes. To second order in the deformation parameter the two equations are the dilaton equation, which recall is independent of the RR sector \cite{Bakhmatov:2018apn}:
\be
K^{\alpha}_i K^{\beta}_k \nabla_{\alpha} K_{\beta m} ( c_{l_1 l_2}^{~~~ m} r^{i l_1} r^{k l_2} + c_{l_1 l_2}^{~~~k} r^{m l_1} r^{i l_2} + c_{l_1 l_2}^{~~~i} r^{k l_1} r^{m l_2} ) = 0,  
\ee
and the Einstein equation, 
\be
\frac{1}{2}  ( {\nabla}_{\rho} K_{ i \mu}  K_{ j \nu} K^{\rho}_{k}  + {\nabla}_{\rho} K_{ i \nu}  K_{ j \mu} K^{\rho}_{k}) \left( c_{l_1 l_2}^{~~~i} r^{j l_1} r^{k l_2} + c_{l_1 l_2}^{~~~k} r^{i l_1} r^{j l_2} + c_{l_1 l_2}^{~~~j} r^{k l_1} r^{i l_2}\right) = \kappa^2 T^{(0)}_{\mu \nu}, 
\ee
where we have added the contribution from the dilaton shift, $\phi \rightarrow  \phi + \log \kappa$, and $T^{(0)}_{\mu \nu}$ is the undeformed stress-energy tensor. Despite not being exact, these perturbative results we can regard as necessary conditions for the existence of a deformation. We omit the $B$-field equation of motion as it recovers the CYBE only at third-order in $\Theta$. 

It is clear from the Einstein equation that the dilaton shift back-reacts once $T^{(0)}_{\mu \nu} \neq 0$. Therefore, we do not have to deform flat directions, however to counteract the effect of the dilaton shift we have to in effect deform components of the geometry that are curved. Since our deformation is only defined for Killing directions, this restricts us to direct-products of coset spaces. In effect this rules out deformations of warped geometries, since the warp factor will depend on a non-Killing direction that cannot be deformed. This restriction is echoed in the dilaton equation since if non-trivial coordinate dependence appeared in this equation, one could no longer trade off the bracketed algebraic terms against one another. Even with cosets, one still has to choose the Killing vectors correctly so that this can be achieved. 

\section{$AdS_3 \times S^3$ revisited}
\label{sec:ads3s3}
Having laid out our stall, in this section we turn to explicit examples and revisit deformations of $AdS_3 \times S^3 \times T^4$ where $\Theta$ is an $r$-matrix solution to the modified CYBE. Our motivation for doing so is two-fold. First, $AdS_3 \times S^3 \times T^4$ is usually supported by three-form flux and not five-form flux, as was the case of the example studied in the previous section \footnote{One can always T-dualise on $T^4$ to replace the three-form with five-form flux. Note, this is not possible for $AdS_3 \times S^3 \times S^3 \times S^1$.}, so one can ask whether this complicates the descent procedure involving the Page forms. 

Secondly, the isometry group of $AdS_2 \times S^2$ is $SL(2, \mathbb{R}) \times SU(2)$, whereas $AdS_3 \times S^3$ has double the symmetry, i. e. $SL(2, \mathbb{R})_{L} \times SL(2, \mathbb{R})_{R} \times SU(2)_L \times SU(2)_R$, so we would like to contemplate separate deformations involving the different factors. In particular, since the geometry is a product of left and right symmetries, it is reasonable to expect that one can have a deformation with two parameters. Such a deformation has appeared previously in the literature \cite{Hoare:2014oua, Lunin:2014tsa}, where it is related to the two-parameter deformation of Fateev \cite{Fateev:1996ea}. To be more precise, the deformed metric was extracted from the Yang-Baxter $\sigma$-model in \cite{Hoare:2014oua} and in \cite{Lunin:2014tsa} a standard supergravity embedding was presented. To the extent of our knowledge a \textit{generalised} supergravity solution including the dilaton, Killing vector $I$ and RR sector has not been presented. This provides a fitting test for our method. 

However, before getting to that point, let us meander a bit to understand the nuts and bolts of the solution generating technique. With a view to isolating the different symmetries, we will employ slightly atypical coordinates and this turns out to be an instructive exercise. Let us begin by recalling the original $AdS_3 \times S^3 \times T^4$ solution, 
\bea
\dd s^2 &=& \dd s^2(AdS_3) + \dd s^2(S^3) + \dd s^2(T^4), \nn
 F_3 &=& 2 \vol (AdS_3) + 2 \vol(S^3),   
\eea
where we will use the following ``Hopf-fibre" metrics for $AdS_3$, 
\bea
\dd s^2(AdS_3) &=& \frac{1}{4} \left( - \dd \tau^2 + \dd \omega^2 + \dd \sigma^2 + 2 \sinh \omega \dd \tau \dd \sigma \right), 
\eea
and $S^3$, 
\bea
\dd s^2(S^3) &=& \frac{1}{4} \left( \dd \theta^2 + \dd \phi^2 + \dd \psi^2 + 2 \cos \theta \dd \phi \dd \psi \right). 
\eea
It is well known that one can analytically continue $AdS_p$ to $S^p$, and vice versa, so that the line elements are related by an overall change in signature $\dd s^2(AdS_p) = - \dd s^2(S^p)$. For the above metrics this can be achieved by identifying coordinates as follows, 
\be
\label{ac1}
\theta = i \omega - \frac{\pi}{2}, \quad \psi = \tau, \quad \phi = i \sigma, 
\ee
in order that one maps the $S^3$ metric to the $AdS_3$ metric with opposite signature.

Given the above parametrisation of the spacetime metric, the six $AdS_3$ Killing vectors can be written as, 
\bea
K_1 &=& - \frac{\sinh \sigma}{\cosh \omega} \partial_{\tau} - \cosh \sigma \partial_{\omega} + \tanh \omega \sinh \sigma \partial_{\sigma}, \nn
K_2 &=&  \frac{ \cosh \sigma}{\cosh \omega} \partial_{\tau} + \sinh \sigma \partial_{\omega} - \tanh \omega \cosh \sigma \partial_{\sigma}, \nn
K_3 &=& \partial_{\sigma}, 
\eea
and
\bea
\tilde{K}_1 &=&  \sin \tau \tanh \omega \partial_{\tau} -  \cos \tau \partial_{\omega} + \frac{\sin \tau}{\cosh \omega} \partial_{\sigma}, \nn
\tilde{K}_2 &=& - \cos \tau \tanh \omega \partial_{\tau} - \sin \tau \partial_{\omega} - \frac{\cos \tau}{\cosh \omega} \partial_{\sigma}, \nn
\tilde{K}_3 &=& \partial_{\tau}. 
\eea
One can check that the two sets of Killing vectors commute with each other and within each set, the $SL(2,\mathbb{R})$ symmetry is manifest through the commutation relations. The corresponding Killing vectors for $S^3$ take the form,  
\bea
K_4 &=& - \frac{\cos \phi}{\sin \theta} \partial_{\psi} + \sin \phi \partial_{\theta} + \cot \theta \cos \phi \partial_{\phi}, \nn
K_5 &=& \frac{\sin \phi}{\sin \theta} \partial_{\psi} + \cos \phi \partial_{\theta} - \cot \theta \sin \phi \partial_{\phi}, \nn
K_6 &=& \partial_{\phi}, 
\eea
and 
\bea
\tilde{K}_4 &=& - \cot \theta \cos \psi \partial_{\psi} - \sin \psi \partial_{\theta} + \frac{\cos \psi}{\sin \theta} \partial_{\phi}, \nn
\tilde{K}_5 &=& - \cot \theta \sin \psi \partial_{\psi} + \cos \psi \partial_{\theta} + \frac{\sin \psi}{\sin \theta} \partial_{\phi}, \nn
\tilde{K}_6 &=& \partial_{\psi}. 
\eea
As may be anticipated, the Killing vectors are not completely independent and they can also be mapped under the analytic continuation (\ref{ac1}):
\be
K_4 = K_2, \quad K_5 = i K_1, \quad K_6 = - i K_3, \quad \tilde{K}_4 = - i \tilde{K}_2, \quad \tilde{K}_5 = i \tilde{K}_1, \quad \tilde{K}_6 = \tilde{K}_3. 
\ee
Having introduced coordinates that make the symmetries manifest and explained the relation between the two metrics under analytic continuation, to define our deformation we need to specify $\Theta$. Making use of the symmetries, the most general $\Theta$ with constant coefficients takes the form,  
\be
\Theta = \alpha K_1 \wedge K_2 + \beta \tilde{K}_1 \wedge \tilde{K}_2 + \gamma K_4 \wedge K_5 + \delta \tilde{K}_4 \wedge \tilde{K}_5, 
\ee
where $\alpha, \beta, \gamma$ and $\delta$ are constant coefficients. In terms of components, $\Theta$ may be expressed as,   
\bea
\Theta^{\tau \omega} &=& \frac{(\alpha - \beta \sinh \omega)}{\cosh \omega}, \quad \Theta^{\omega \sigma} = \frac{(\alpha \sinh \omega + \beta)}{\cosh \omega}, \nn
\Theta^{\theta \phi} &=& - \frac{(\delta + \gamma \cos \theta)}{\sin \theta}, \quad \Theta^{\psi \theta} = - \frac{( \delta \cos\theta + \gamma)}{\sin \theta}, 
\eea
where, due to our conveniently chosen Killing vectors, $\Theta$ only depends on $\omega$ and $\theta$. 

If we had chosen different Killing vectors, which are equivalent up to $SL(2, \mathbb{R})$ or $SU(2)$ transformations, dependence on the remaining coordinates would have crept in. For this reason, this is the simplest $\Theta$, but also the most general modulo symmetry transformations. 

With $\Theta$ specified, it is a straightforward exercise to exploit (\ref{open_closed}), (\ref{I_theta}) and (\ref{tduality_inv}) to write out the deformed metric,  
\bea
\label{deformed_su2_su2}
\dd s^2 &=& - \frac{(16+ \beta^2 + 2 \alpha \beta \sinh \omega + \alpha^2 \sinh^2 \omega)}{4 \Delta_1} \dd \tau^2 + \frac{(16 - \alpha^2 + 2 \alpha \beta \sinh \omega - \beta^2 \sinh^2 \omega)}{4 \Delta_1} \dd \sigma^2 \nn
&+& \frac{4}{\Delta_1} \dd \omega^2 + \frac{((16-\alpha^2 + \beta^2) \sinh \omega + \alpha \beta \sinh^2 \omega - \alpha \beta)}{2 \Delta_1} \dd \tau \dd \sigma \nn
&+&  \frac{(16 + \gamma^2 + 2 \gamma \delta \cos \theta + \delta^2 \cos^2 \theta)}{4 \Delta_2} \dd \phi^2 + \frac{(16 + \delta^2 + 2 \gamma \delta \cos \theta + \gamma^2 \cos^2 \theta)}{4 \Delta_2} \dd \psi^2 \nn
&+& \frac{4}{\Delta_2} \dd \theta^2 + \frac{(\gamma \delta + (16 + \gamma^2 + \delta^2) \cos \theta + \gamma \delta \cos^2 \theta)}{2 \Delta_2} \dd \phi \dd \psi, \nonumber
\eea
and supporting fields from the NS sector, 
\bea
B &=&  - \frac{\cosh \omega}{\Delta_1} \dd \omega \wedge (  \alpha \dd \tau  + \beta \dd \sigma) + \frac{\sin \theta}{\Delta_2} \dd \theta \wedge ( \delta \dd \phi - \gamma \dd \psi ), \nn
 \phi &=& - \frac{1}{2} \log \left[ \frac{\Delta_1 \Delta_2}{256} \right] + \phi_0, \nn
 I &=& \beta \partial_{\tau}  + \alpha \partial_{\sigma} +  \gamma \partial_{\phi} - \delta \partial_{\psi}. 
\eea
Here we have introduced, 
\be
\Delta_1 = 16 - \alpha^2 + \beta^2 + 2 \alpha \beta \sinh \omega, \quad \Delta_2 = 16 + \gamma^2 + \delta^2 + 2 \gamma \delta \cos \theta. 
\ee
To accommodate deformations based on the modified CYBE we have also allowed for a constant shift in the dilaton $\phi_0$. It is easy to check that the $B$-field is closed, i. e. $H=0$, so that our examination of the equations of motion will be somewhat simplified.  

With the deformed NS sector in hand, bearing in mind that the dilaton equation (\ref{general3}) only involves fields from this sector, we are now in a position to document our first constraint on the coefficients: 
\be
\label{dilaton_constraint}
\alpha^4 + \beta^4 + 96 \beta^2 - 96 \alpha^2 + 2 \alpha^2 \beta^2 - \gamma^4 - 96 \gamma^2 + 2 \gamma^2 \delta^2 - 96 \delta^2 - \delta^4 = 0. 
\ee

To go further and specify the complete solution, we identify the invariant Page forms. In contrast to geometries supported by five-form flux, where there is only a Page five-form, here we have a three-form flux, so we can define both three and seven-form Page forms:
\bea
Q_3 &=& \frac{1}{4} \cosh \omega \dd \tau \wedge \dd \omega \wedge \dd \sigma + \frac{1}{4} \sin \theta \dd \theta \wedge \dd \phi \wedge \dd \psi, \nn
Q_7 &=& - Q_3 \wedge \vol(T^4). 
%Q_7 &=& -  \left[ \frac{1}{4} \cosh \omega \dd \tau \wedge \dd \omega \wedge \dd \sigma+ \frac{1}{4} \sin \theta \dd \theta \wedge \dd \phi \wedge \dd \psi \right] \wedge \vol(T^4).  
\eea
The presence of two Page forms simply reflects the fact that the geometry is sourced by D1 and D5-branes and not intersecting D3-branes and this marks a departure from the earlier example. Regardless, our descent procedure still works and employing it we can define an induced one-form and a five-form: 
\bea
Q_1 &=& - \frac{1}{4} ( \alpha - \beta \sinh \omega) \dd \sigma  - \frac{1}{4} ( \alpha \sinh \omega + \beta)  \dd \tau + \frac{1}{4} (\delta + \gamma \cos \theta)  \dd \psi \nn 
&+& \frac{1}{4} ( \delta \cos \theta + \gamma)  \dd \phi, \quad Q_5 = - Q_1 \wedge \vol (T^4). 
%Q_5 &=& \frac{1}{4} ( \alpha - \beta \sinh \omega) \dd \sigma \wedge \vol(T^4) + \frac{1}{4}  ( \alpha \sinh \omega + \beta)  \dd \tau \wedge \vol(T^4) \nn
%&-& \frac{1}{4} ( \delta + \gamma \cos \theta) \dd \psi \wedge \vol(T^4) - \frac{1}{4} ( \delta \cos \theta + \gamma ) \dd \phi \wedge \vol(T^4). 
\eea
The one-form $Q_1$ is induced from $Q_3$, while the five-form $Q_5$ is induced from $Q_7$. It is straightforward to check that $\dd Q_1 = i_{I} Q_3$, which implies that $\dd Q_5 = i_{I} Q_7$. This leaves only the $\dd Q_3 = i_{I} Q_5$ equation of motion from (\ref{dQQ}), which is not obviously satisfied by our descent procedure since there is no induced component to $Q_3$, which means that it must be closed. As a result, we identify a further constraint,   
\bea
\label{Q5_constraint}
i_{I} Q_5 &=& \left[ \frac{\beta}{4} ( \alpha \sinh \omega + \beta) + \frac{\alpha}{4} (\alpha - \beta \sinh \omega) + \frac{\delta}{4} ( \delta + \gamma \cos \theta) - \frac{\gamma}{4} ( \delta \cos \theta + \gamma ) \right]  \vol(T^4), \nn
&=& \frac{1}{4} \left( \alpha^2 + \beta^2 + \delta^2 - \gamma^2 \right) \vol(T^4) = 0. 
\eea
The above signs are interesting, since assuming everything is real, there is no way to identify constants in a symmetric fashion consistent with analytic continuation (\ref{ac1}). This suggests that at least one of the constants is pure imaginary. To better understand this point, we recall that the metrics are related by analytic continuation, which in turn relates the constants in the following manner:
\be
\alpha = - i \gamma, \quad \beta = - \delta. 
\ee
One immediate consequence of imposing these conditions is that the constraint coming from the dilaton equation (\ref{dilaton_constraint}) is trivially satisfied. Furthermore, the constraint (\ref{Q5_constraint}) can be satisfied for two constants $\kappa_{\pm}$, 
\be
\label{branches}
- \beta = \delta = \pm \gamma = \pm i \alpha = 2 \kappa_{\pm}, 
\ee
which may be distinguished by a choice of signs. At this juncture, we should be a little concerned: assuming the remaining equations of motion hold and a solution exists, it is clear that the final geometry will be complex. In contrast, the solutions quoted in the literature, e. g. \cite{Arutyunov:2015mqj}, are all real. That being said, the known solutions differ in an obvious way: the coordinates are not the same. To reconcile everything, it is reasonable to suspect there exists a coordinate transformation, essentially another analytic continuation, through which a real solution is produced. This is indeed the case, as we now explain. 

To get a real solution, one must entertain the following coordinate transformations: 
\bea
\sin \frac{\theta}{2} &=& r, \quad  \psi = \varphi + \phi_1, \quad  \phi = \varphi - \phi_1, \nn 
\sin ( \frac{i \omega}{2} - \frac{\pi}{4} ) &=& i \rho, \quad \tau = t + \psi_1, \quad \sigma = - i ( t - \psi_1). 
\eea

We are now in a position to analyse the two branches (\ref{branches}) to recover known solutions from the literature.  Let us start with the $\kappa_-$ deformation, which will turn out to be more familiar, since it has been quoted more often in the literature. Employing the coordinate transformation, the deformation parameter may be expressed as, 
\be
\label{deform_minus}
\Theta_-^{t \rho} = \kappa_- \rho, \quad \Theta_{-}^{\varphi r} = \kappa_- r, 
\ee
and the NS sector for the solution becomes: 
\bea
\dd s^2 &=& - \frac{(1+\rho^2)}{(1-\kappa_-^2 \rho^2)} \dd t^2 + \frac{ \dd \rho^2}{(1 - \kappa_-^2 \rho^2) (1 + \rho^2)} + \rho^2 \dd \psi_1^2 \nn
&+& \frac{(1-r^2)}{(1 + \kappa_-^2 r^2)} \dd \varphi^2 + \frac{\dd r^2}{(1+ \kappa_-^2 r^2)(1-r^2)}+ r^2 \dd \phi_1^2 + \dd s^2 (T^4), \nn
B &=& \frac{\kappa_- \rho}{(1 - \kappa_-^2 \rho^2)} \dd t \wedge \dd \rho - \frac{\kappa_- r}{(1+ \kappa_-^2 r^2)} \dd \varphi \wedge \dd r, \nn
\phi &=& - \frac{1}{2} \log [ ( 1 - \kappa_-^2 \rho^2)(1 + \kappa_-^2 r^2) ] + \phi_0, \quad I = - 2 \kappa_- ( \partial_t + \partial_{\varphi}),  
\eea
where we have retained the constant shift in the dilaton. Neglecting $Q_7$, which is simply the Hodge dual of $Q_3$, the key Page forms are 
\bea
Q_1 &=& 2 \kappa_-( - \rho^2 \dd \psi_1 + r^2 \dd \phi_1), \nn
Q_3 &=& 2 \rho \dd t \wedge \dd \rho \wedge \dd \psi_1 - 2 r \dd \varphi \wedge \dd r \wedge \dd \phi_1, \nn
Q_5 &=& 2 \kappa_- ( \rho^2 \dd \psi_1 - r^2 \dd \phi_1) \wedge \vol(T^4), 
\eea
which can be unravelled to extract the RR field strengths: 
\bea
F_1 &=& 2 \kappa_-( - \rho^2 \dd \psi_1 + r^2 \dd \phi_1), \nn
F_3 &=&  \frac{2 \rho}{(1- \kappa_-^2 \rho^2)} \dd t \wedge \dd \rho \wedge \dd \psi_1 - \frac{2 r}{(1+ \kappa_-^2 r^2)} \dd \varphi \wedge \dd r \wedge \dd \phi_1
- \frac{2 \kappa_-^2 \rho r^2}{(1-\kappa_-^2 \rho^2)} \dd t \wedge \dd \rho \wedge \dd \phi_1 \nn &-& \frac{2 \kappa_-^2 r \rho^2}{(1+\kappa_-^2 r^2)} \dd \varphi \wedge \dd r \wedge \dd \psi_1, \nn
F_5 &=& \frac{2 \kappa_- r \rho}{(1-\kappa_-^2 \rho^2)(1 + \kappa_-^2 r^2)} \dd t \wedge \dd \rho \wedge \dd \varphi \wedge \dd r \wedge (\dd \phi_1 + \dd \psi_1) \nn
&+& 2 \kappa_- ( \rho^2 \dd \psi_1 - r^2 \dd \phi_1) \wedge \vol(T^4). 
\eea
To complete the solution, one needs to fix the constant shift in the dilaton and this can be done using either the Einstein (\ref{general1}) or the $B$-field equation of motion (\ref{general2}):  
\be
e^{2 \phi_0} = 1 + \kappa_-^2. 
\ee

Let us now return to the $\kappa_+$ branch. In this case the deformation parameter is 
\be
\label{deform_plus}
\Theta_{+}^{\rho \psi_1} = - \kappa_+ \left( \frac{1}{\rho} + \rho \right), \quad \Theta_{+}^{r \phi_1} =  \kappa_+ \left( \frac{1}{r} - r \right), 
\ee
and once re-written in the new coordinates, the deformed NS sector takes the form: 
\bea
\dd s^2 &=& - (1+\rho^2) \dd t^2 + \frac{ \dd \rho^2}{(1 +\kappa_+^2 (1 + \rho^2)) (1 + \rho^2)} + \frac{\rho^2}{(1 + \kappa_+^2 (1 + \rho^2))} \dd \psi_1^2 \nn
&+&  ( 1 - r^2) \dd \varphi^2 +\frac{\dd r^2}{(1-r^2)( 1+ \kappa_+^2 (1-r^2))} +  \frac{r^2  \dd \phi_1^2}{(1 + \kappa_+^2 (1-r^2))} + \dd s^2 (T^4), \nn
B &=& \frac{\kappa_+ \rho}{(1 + \kappa_+^2 ( 1 +\rho^2) )} \dd \rho \wedge \dd \psi_1 - \frac{\kappa_+ r}{(1+ \kappa_+^2 (1-r^2))} \dd r \wedge \dd \phi_1, \nn
\phi &=& - \frac{1}{2} \log [ ( 1 + \kappa_+^2 (1+\rho^2))(1 + \kappa_+^2(1- r^2)) ] + \phi_0, \quad I = - 2 \kappa_+ ( \partial_{\psi_1} + \partial_{\phi_1}).  
\eea
We omit further expressions, but simply remark that one can analytically continue the two solutions into each other starting with the $\kappa_-$ deformation. The required analytic continuation is  
\be
\rho \rightarrow i \sqrt{1+ \rho^2}, \quad t \rightarrow \psi_1, \quad r \rightarrow \sqrt{1-r^2}, \quad \varphi \rightarrow \phi_1, \quad \kappa_- \rightarrow \kappa_+.   
\ee

A number of comments are in order. The simplest observation is that even though we have used unusual coordinates, our prescription works for $AdS_3 \times S^3$ and that the RR sector can be constructed via descent from the Page forms. While we have taken pains to isolate the different symmetry factors and their respective Killing vectors, we see that the reality, or otherwise, of the final solution may depend on the choice of coordinates, or alternatively, the Killing vectors. We noted that the deformed solution was complex in the original coordinates we adopted, but through analytic continuation, it could be made real \footnote{This is somewhat reminiscent of $\lambda$-deformations of $AdS_3 \times S^3$ based on $SL(2, \mathbb{R})$ and $SU(2)$ group manifolds \cite{Sfetsos:2014cea}, but there it is simply the RR sector that is complex, whereas here the spacetime metric is also complex in the original coordinates.}. The fact that we arrive at a complex deformation is expected to be an artifact of the fact that we have also imposed a requirement that the deformations respect the analytic continuation, which clearly complexifies Killing vectors. Presumably if one relaxes this condition, an independent real solution can be found in the original coordinates. Note, even in global coordinates the underlying Killing vectors are complex (see appendix B of \cite{Araujo:2017jap}), but they conspire to give a real deformation parameter $\Theta$, so the final geometry is real. 

\subsection{Two-parameter q-deformation}
Now that we have recovered the known solutions, we can consider the two-parameter q-deformation \cite{Hoare:2014oua, Lunin:2014tsa}. In \cite{Hoare:2014oua} only the metric appears, while in \cite{Lunin:2014tsa}, the metric is embedded in type II supergravity, where the solution is supported by an RR three-form and one-form. Here, we provide a direct embedding in generalised supergravity, where as is now standard in our approach, we directly use (\ref{I_theta}) to specify the Killing vector. 

In order to write down the solution at this stage all we have to do is simply combine the earlier deformations (\ref{deform_minus}) and (\ref{deform_plus}) and use the map (\ref{open_closed}). Doing so, the NS sector becomes, 
\bea
\dd s^2 &=& \frac{1}{X_1} \biggl[ - ( 1+ \kappa_+^2 (1+ \rho^2))(1 + \rho^2) \dd t^2 +  (1 - \kappa_-^2 \rho^2) \rho^2 \dd \psi_1^2 + \frac{\dd \rho^2}{(1+ \rho^2)}  \nn
&+&  2 \kappa_+ \kappa_- \rho^2 (1 + \rho^2) \dd t \dd \psi_1 \biggr] + \frac{1}{X_2} \biggl[ (1 + (1-r^2) \kappa_+^2) (1-r^2) \dd \varphi^2  + (1 + \kappa_-^2 r^2) r^2 \dd \phi_1^2 \nn 
&+& \frac{\dd r^2}{(1-r^2)} + 2 \kappa_+ \kappa_- r^2 (1 -r^2) \dd \varphi \dd \phi_1 \biggr]  + \dd s^2 (T^4), \nn
B &=& - \frac{\rho \dd \rho}{X_1} \wedge ( \kappa_- \dd t - \kappa_+ \dd \psi_1) + \frac{r \dd r}{X_2} \wedge ( \kappa_- \dd \varphi - \kappa_+ \dd \phi_1), \nn
\phi &=& - \frac{1}{2} \log ( X_1 X_2) + \phi_0, \quad I = - 2 \left( \kappa_- \partial_t + \kappa_+ \partial_{\psi_1} + \kappa_- \partial_{\varphi} + \kappa_+ \partial_{\phi_1}\right), 
\eea
where we have defined, 
\be
X_1 = 1- \kappa_-^2 \rho^2 + \kappa_+^2 (1+ \rho^2), \quad X_2 = 1 + \kappa_+^2 (1-r^2)+ \kappa_-^2 r^2. 
\ee
One can check that the dilaton equation (\ref{general3}) is satisfied, so we are clearly on the right track, since one of the non-trivial equations holds. 

Using our descent procedure for the RR sector, the deformed Page forms become: 
\bea
Q_1 &=& 2 \kappa_+( (1+\rho^2) \dd t + (1-r^2) \dd \varphi) + 2 \kappa_-( - \rho^2 \dd \psi_1 + r^2 \dd \phi_1), \nn
Q_3 &=& 2 \rho \dd t \wedge \dd \rho \wedge \dd \psi_1 - 2 r \dd \varphi \wedge \dd r \wedge \dd \phi_1, \quad Q_5 = - Q_1 \wedge \vol(T^4). 
%Q_5 &=& \left[-2 \kappa_+ ( (1+\rho^2) \dd t + (1-r^2) \dd \varphi) + 2 \kappa_- ( \rho^2 \dd \psi_1 - r^2 \dd \phi_1) \right] \wedge \vol(T^4). 
\eea
Unravelling these expressions, we find the RR field strengths: 
\bea
F_1 &=& 2 \kappa_+( (1+\rho^2) \dd t + (1-r^2) \dd \varphi) + 2 \kappa_-( - \rho^2 \dd \psi_1 + r^2 \dd \phi_1), \nn
F_3 &=& \frac{1}{X_1} \biggl[ 2 \rho \dd t \dd \rho \dd \psi_1 - 2 \kappa_+ \kappa_- \rho (1-r^2) \dd t \dd \rho \dd \varphi - 2 \kappa_-^2 r^2 \rho \dd t \dd \rho \dd \phi_1 - 2 \kappa_+^2 \rho (1-r^2) \dd \rho \dd \psi_1 \dd \varphi \nn 
&-& 2 \kappa_+ \kappa_- \rho r^2 \dd \rho \dd \psi_1 \dd \phi_1  \biggr] + \frac{1}{X_2} \biggl[ - 2 r \dd \varphi \dd r \dd \phi_1 - 2 \kappa_+ \kappa_- r (1+\rho^2) \dd r \dd \varphi \dd t + 2 \kappa_-^2 r \rho^2 \dd r \dd \varphi \dd \psi_1 \nn 
&+& 2 \kappa_+^2 r (1+ \rho^2) \dd r \dd \phi_1 \dd t - 2 \kappa_+ \kappa_- r \rho^2 \dd r \dd \phi_1 \dd \psi_1 \biggr], \nn
F_5 &=&  \frac{2 \rho r}{X_1 X_2} \biggl[ \kappa_- \dd t \dd \rho \dd \psi_1 \dd \varphi \dd r + \kappa_- \dd t \dd \rho \dd \varphi \dd r \dd \phi_1+  \kappa_+  \dd t \dd \rho \dd \psi_1 \dd r \dd \phi_1 \nn 
&+&  \kappa_+ \dd \rho \dd \psi_1 \dd \varphi \dd r \dd \phi_1 \biggr] \nn
&+& \left[-2 \kappa_+ ( (1+\rho^2) \dd t + (1-r^2) \dd \varphi) + 2 \kappa_- ( \rho^2 \dd \psi_1 - r^2 \dd \phi_1) \right] \wedge \vol(T^4). 
\eea

It can be checked that the remaining equations of motion are satisfied provided, 
\be
e^{2 \phi_0} = (1 + \kappa_+^2)(1+ \kappa_-^2). 
\ee

Our analysis shows that there is a two-parameter q-deformation of $AdS_3 \times S^3$ giving rise to a solution to generalised supergravity. In essence, we have simply written it down using the methods outlined in section \ref{sec:method}. The existence of this solution was anticipated in earlier work \cite{Hoare:2014oua, Lunin:2014tsa}, but to our knowledge never completed. 

\section{$AdS_3 \times S^3 \times S^3 \times S^1$}
\label{sec:ads3s3s3}
Having discussed the simpler setting of $AdS_3 \times S^3$, let us move onto $AdS_3 \times S^3 \times S^3$, which exhibits large superconformal symmetry and comprises two copies of the exceptional Lie supergroup $D(2, 1;  \alpha)$. From the perspective of geometry, we recall that the constant $\alpha$ is the ratio between the radii of the three-spheres and it is our goal here to check that deformations exist for all values of $\alpha$ and construct the solution explicitly. To avoid added complexity, and the resulting lengthy expressions, we will simply deform one of these $D(2, 1;  \alpha)$ factors with bosonic subgroup $SL(2, \mathbb{R}) \times SU(2) \times SU(2)$. The generalisation to both copies of $D(2, 1; \alpha)$ is straightforward and follows similar lines to the previous section. 

Before proceeding to the analysis, let us advertise one interesting feature of the deformation. In contrast to the previous example where there was no induced Page three-form $Q_3$, here we will find that it is induced from the (closed) Page seven-form $Q_7$. This means that there are two contributions to $Q_3$: one is inherited from the original geometry, while the other is induced. This is a feature of more complicated deformations. For example, it is relatively straightforward to extend our analysis here to IIA and cosets, such as $\AdS_4 \times M_6$, where $M_6$ is $\mathbb{C} P^3$, $\mathbb{C} P^2 \times S^2$, $S^2 \times S^2 \times S^2$, etc. \footnote{See \cite{Duff:1986hr} for a review.}. In the process, it is clear one will generate examples of deformed geometries supported by a zoo of RR field strengths. 

But back to the task at hand. We recall that the undeformed geometry may be written as,  
\bea
\dd s^2 &=& L^2 \dd s^2(AdS_3) + R_1^2 \dd s^2 (S^3_1) + R_2^2 \dd s^2 (S^3_2) + \dd x^2, \nn
%- (1 + \rho^2) \dd t^2 + \frac{\dd \rho^2}{(1+ \rho^2)} + \rho^2 \dd \psi^2  + \sum_{i=1}^2 \left[ (1-r_i^2) \dd \varphi_i^2 + \frac{\dd r_i^2}{(1-r_i^2)} + r_i^2 \dd \phi_i^2 \right]  + \dd x^2, \nn
F_3 &=& 2 L^2 \vol (AdS_3) + 2 R_1^2 \vol (S^3_1) + 2 R_2^2 \vol (S^3_2), 
\eea
where adopting the lesson from the last section, we will employ global coordinates, 
\bea
\dd s^2 (AdS_3) &=& - (1 + \rho^2) \dd t^2 + \frac{\dd \rho^2}{(1+ \rho^2)} + \rho^2 \dd \psi^2, \nn
\dd s^2 (S^3_i) &=& (1-r_i^2) \dd \varphi_i^2 + \frac{\dd r_i^2}{(1-r_i^2)} + r_i^2 \dd \phi_i^2, \nn
\vol(AdS_3) &=& \rho \dd t \wedge \dd \rho \wedge \dd \psi, \quad \vol(S^3_i) = r_i \dd \varphi_i \wedge \dd r_i \wedge \dd \phi_i. 
\eea
This is a solution to type IIB supergravity provided the radii satisfy the relation 
\be
\label{radii_constraint}
\frac{1}{L^2} = \frac{1}{R_1^2} + \frac{1}{R_2^2}. 
\ee
As we have seen in the last section, if one neglects the second three-sphere, there is a deformation where the bivector takes the form (\ref{deform_minus}). This suggests that we can extend the deformation to the second three-sphere by simply repeating it, 
\be
\Theta^{t \rho} = \kappa_0 \rho, \quad \Theta^{\varphi_1 r_1} = \kappa_1 r_1, \quad \Theta^{\varphi_2 r_2} = \kappa_2 r_2. 
\ee
Since we have yet to identify the relation between the deformation parameters, for the moment $\kappa_0, \kappa_1$ and $\kappa_2$ are arbitrary constants. The plan now is to use our prescription to fix them and check that the constant $\alpha$ drops, or put alternatively, that there is a choice of constants so that the constraint on the radii (\ref{radii_constraint}) still holds. 

First, let us perform the transformation (\ref{open_closed}) with a view to identifying the deformed metric and $B$-field. The resulting expressions are, 
\bea
g_{\mu \nu} \dd x^{\mu} \dd x^{\nu} &=& - \frac{L^2 (1 + \rho^2)}{(1 - L^4 \kappa_0^2 \rho^2)} \dd t^2 + \frac{L^2}{(1+\rho^2)(1-L^4 \kappa_0^2 \rho^2)} \dd \rho^2 + L^2 \rho^2 \dd \psi^2 \nn
&+& \frac{R_1^2 (1-r_1^2) }{(1 + R_1^4 \kappa_1^2 r_1^2)} \dd \varphi_1^2 + \frac{R_1^2}{(1-r_1^2)(1 + R_1^4 \kappa_1^2 r_1^2)} \dd r_1^2 + R_1^2 r_1^2 \dd \phi_1^2 \nn 
&+& \frac{R_2^2 (1-r_2^2) }{(1 + R_2^4 \kappa_2^2 r_2^2)} \dd \varphi_2^2 + \frac{R_2^2}{(1-r_2^2)(1 + R_2^4 \kappa_2^2 r_2^2)} \dd r_2^2 + R_2^2 r_2^2 \dd \phi_2^2 + \dd x^2, \\
B &=& \frac{L^4 \kappa_0 \rho}{(1 - L^4 \kappa_0^2 \rho^2)} \dd t \wedge \dd \rho  - \frac{R_1^4 r_1 \kappa_1}{(1 + R_1^4 \kappa_1^2 r_1^2)} \dd \varphi_1 \wedge \dd r_1 - \frac{R_2^4 r_2 \kappa_2}{(1 + R_2^4 \kappa_2^2 r_2^2)} \dd \varphi_2 \wedge \dd r_2. \nonumber
\eea
It should be noted that $H = \dd B = 0$ and it will not feature in the equations of motion. 

From (\ref{I_theta}) we can determine the Killing vector, 
\be
I = - 2 ( \kappa_0 \partial_t + \kappa_1 \partial_{\varphi_1} + \kappa_2 \partial_{\varphi_2}), 
\ee
while the dilaton, modulo the usual constant $\phi_0$, follows from the T-duality invariant (\ref{tduality_inv}): 
\be
e^{-2 \phi} = e^{2 \phi_0}(1 - L^4 \kappa_0^2 \rho^2) (1+ R_1^4 \kappa_1^2 r_1^2) (1+ R_2^4 \kappa_2^2 r_2^2). 
\ee
 
At this juncture, we can once again go ahead and check the dilaton equation as it does not involve the RR sector. The dilaton equation is satisfied provided, 
\be
\label{dilaton_EOM}
- \frac{1}{L^2} + \frac{1}{R_1^2} + \frac{1}{R_2^2} + L^2 \kappa_0^2  - R_1^2 \kappa_1^2 - R_2^2 \kappa_2^2 = 0. 
\ee
This is the first indication that there is a good deformation. The $\kappa$-independent terms vanish using the constraint on the radii, as expected, leaving a constraint on the constants to be imposed. 

To make use of our procedure, we identify the original Page forms, 
\bea
Q_3 &=& 2 L^2 \vol (AdS_3) + 2 R_1^2 \vol (S^3_1) + 2 R_2^2 \vol (S^3_2), \nn
Q_7 &=& - \frac{2 R_1^3 R_2^3}{L} \vol(S_1^3) \wedge \vol(S_2^3) \wedge \dd x - \frac{2 L^3 R_2^3}{R_1} \vol(AdS_3) \wedge \vol(S_2^3) \wedge \dd x \nn
&+& \frac{2 L^3 R_1^3}{R_2} \vol (AdS_3) \wedge \vol(S_1^3) \wedge \dd x, 
\eea
and define induced Page forms by descent, starting with the one-form $Q_1$, which can be unambiguously identified, 
\be
Q_{1\, \rho} = - (\Theta \lrcorner Q_3)_\rho \equiv - \frac{1}{2} \Theta^{\mu \nu} Q_{3 \, \mu \nu \rho}.  
\ee
The sign is essentially fixed by the requirement that $ \dd Q_1 = i_{I} Q_3$. This allows us to identify the one-form $F_1$, 
\bea
F_1 &=& - 2 L^2 \kappa_0 \rho^2 \dd \psi - 2 R_1^2 \kappa_1 r_1^2 \dd \phi_1 - 2 R_2^2 \kappa_2 r_2^2 \dd \phi_2. 
%&=& - 2 L \kappa_0 \rho e^{\psi} - 2 R_1 \kappa_1 r_1 e^{\phi_1} - 2 R_2 \kappa_2 r_2 e^{\phi_2}, 
\eea

From here we can move onto extracting $F_3$. While there is a contribution to $Q_3$ that is invariant and therefore closed, there is also an additional induced contribution from the Page seven-form, 
\be
\label{Q3Q7}
Q_{3 \, \mu \nu \rho} =  \frac{1}{8} \Theta^{\sigma_1 \sigma_2} \Theta^{\sigma_3 \sigma_4} Q_{7 \, \sigma_1 \sigma_2 \sigma_3 \sigma_4 \mu \nu \rho}. 
\ee

Combining this with the original $Q_3$, we can evaluate $F_3$:
\bea
F_3 &=& Q_3 - B \wedge F_1, \\
&=&  \frac{2 L^2}{(1- L^4 \kappa_0^2 \rho^2)} \left(  \rho \dd t \dd \rho \dd \psi + L^2 R_1^2 \kappa_0 \kappa_1 r_1^2 \rho \dd t \dd \rho \dd \phi_1 + L^2 R_2^2 \kappa_0 \kappa_2 r_2^2 \rho \dd t \dd \rho \dd \phi_2 \right) \nn
&+& \frac{2 R_1^2}{(1 + R_1^4 \kappa_1^2 r_1^2)} \left( r_1 \dd \varphi_1 \dd r_1 \dd \phi_1 -  R_1^2 L^2 \kappa_1 \kappa_0 r_1 \rho^2 \dd \varphi_1 \dd r_1 \dd \psi - R_1^2 R_2^2 \kappa_1 \kappa_2 r_1 r_2^2 \dd \varphi_1 \dd r_1 \dd \phi_2 \right) \nn
&+& \frac{2 R_2^2}{(1 + R_2^4 \kappa_2^2 r_2^2)} \left( r_2 \dd \varphi_2 \dd r_2 \dd \phi_2 -  R_2^2 L^2 \kappa_2 \kappa_0 r_2 \rho^2 \dd \varphi_2 \dd r_2 \dd \psi - R_1^2 R_2^2 \kappa_1 \kappa_2 r_2 r_1^2 \dd \varphi_2 \dd r_2 \dd \phi_1 \right) \nn
&-& \frac{2 R_1^3 R_2^3 \kappa_1 r_1^2 \kappa_2 r_2^2}{L} \dd \phi_1 \dd \phi_2 \dd x  - \frac{2 L^3 R_2^3 \kappa_0 \rho^2 \kappa_2 r_2^2}{R_1} \dd \psi \dd \phi_2 \dd x + \frac{2 L^3 R_1^3 \kappa_0 \rho^2 \kappa_1 r_1^2}{R_2}  \dd \psi \dd \phi_1 \dd x,   \nonumber
\eea
where we have omitted wedge products to save space. Following our prescription, $Q_5$ follows from $Q_7$, again through contraction, 
\bea
Q_{5} &=& - \Theta \lrcorner Q_7, \nn
&=& \frac{2 L^3 R_2^3 \kappa_0 \rho^2}{R_1}  \dd \psi \wedge \vol(S_2^3) \wedge \dd x - \frac{2 L^3 R_1^3 \kappa_0 \rho^2}{R_2} \dd \psi \wedge \vol(S_1^3) \wedge \dd x, \nn
&+& \frac{2 R_1^3 R_2^3 \kappa_1 r_1^2}{L} \dd \phi_1 \wedge \vol(S_2^3) \wedge \dd x + \frac{2 L^3 R_1^3 \kappa_1 r_1^2}{R_2} \dd \phi_1 \wedge \vol(AdS_3) \wedge \dd x, \nn
&-& \frac{2 R_1^3 R_2^3 \kappa_2 r_2^2}{L} \dd \phi_2 \wedge \vol (S^3_1) \wedge \dd x - \frac{2 L^3 R_2^3 \kappa_2 r_2^2}{R_1} \dd \phi_2 \wedge \vol(AdS_3) \wedge \dd x.  
\eea
One can again check that the equations of motion  $ \dd Q_3 = i_{I} Q_5$ and $\dd Q_5 = i_{I} Q_7$ are satisfied. We also require that $ \dd Q_3 = i_{I} Q_5$, otherwise the equations of motion will not be satisfied. This turns out to be the case. 

We are now in a position to identify the lengthy five-form flux,
\bea
F_5 &=& Q_5 - B \wedge F_3 - \frac{1}{2} B^2 \wedge F_1, \\
&=& \frac{2 \kappa_0 L^3 R_2^3 \rho^2 r_2}{R_1 (1 + R_2^4 \kappa_2^2 r_2^2)} \dd \varphi_2 \dd r_2 \dd \psi \dd \phi_2 \dd x - \frac{2 R_1^2 L^4 \kappa_0 r_1 \rho}{(1-L^4 \kappa_0^2 \rho^2)(1+R_1^4 \kappa_1^2 r_1^2)} \dd t \dd \rho \dd \varphi_1 \dd r_1 \dd \phi_1, \nn
&-& \frac{2 \kappa_0 L^3 R_1^3  \rho^2 r_1}{R_2 (1+R_1^4 \kappa_1^2 r_1^2)} \dd \varphi_1 \dd r_1 \dd \psi \dd \phi_1 \dd x  - \frac{2 R_2^2 L^4 \kappa_0 r_2 \rho}{(1-L^4 \kappa_0^2 \rho^2)(1+R_2^4 \kappa_2^2 r_2^2)} \dd t \dd \rho \dd \varphi_2 \dd r_2 \dd \phi_2 \nn
&-& \frac{2 \kappa_1 R_1^3 R_2^3  r_1^2 r_2}{L (1+R_2^4 \kappa_2^2 r_2^2)} \dd \varphi_2 \dd r_2 \dd \phi_2 \dd \phi_1 \dd x  + \frac{2 L^2 R_1^4 \kappa_1 r_1 \rho}{(1-L^4 \kappa_0^2 \rho^2)(1+R_1^4 \kappa_1^2 r_1^2)} \dd t \dd \rho \dd \psi \dd \varphi_1 \dd r_1 \nn
&-& \frac{2 \kappa_1 L^3 R_1^3  r_1^2 \rho}{R_2 (1 - L^4 \kappa_0^2 \rho^2)} \dd t \dd \rho \dd \psi \dd \phi_1 \dd x  + \frac{2 R_1^4 R_2^2 \kappa_1 r_1 r_2}{(1+R_1^4 \kappa_1^2 r_1^2)(1+R_2^4 \kappa_2^2 r_2^2)}  \dd \varphi_2 \dd r_2 \dd \phi_2 \dd \varphi_1 \dd r_1 \nn
&+& \frac{2 \kappa_2 R_1^3 R_2^3 r_2^2 r_1}{L (1+R_1^4 \kappa_1^2 r_1^2)} \dd \varphi_1 \dd r_1 \dd \phi_1 \dd \phi_2 \dd x + \frac{2 R_2^4 L^2 \kappa_2 r_2 \rho}{(1-L^4 \kappa_0^2 \rho^2)(1+R_2^4 \kappa_2^2 r_2^2)} \dd t \dd \rho \dd \psi \dd \varphi_2 \dd r_2  \nn
&+& \frac{2 \kappa_2 L^3 R_2^3 r_2^2 \rho}{R_1(1-L^4 \kappa_0^2 \rho^2)} \dd t \dd \rho \dd \psi \dd \phi_2 \dd x + \frac{2 R_2^4 R_1^2 \kappa_2 r_2 r_1}{(1+R_1^4 \kappa_1^2 r_1^2)(1+R_2^4 \kappa_2^2 r_2^2)} \dd \varphi_1 \dd r_1 \dd \phi_1 \dd \varphi_2 \dd r_2 \nn 
&+& \frac{2 L^4 R_1^4 R_2^2 r_1 \rho r_2^2 \kappa_0 \kappa_1 \kappa_2}{(1- L^4 \kappa_0^2 \rho^2)(1+R_1^4 \kappa_1^2 r_1^2 ) }  \dd t \dd \rho \dd \varphi_1 \dd r_1 \dd \phi_2 + \frac{2 L^3 R_1^3 R_2^3 \rho^2 r_1^2 r_2 \kappa_0 \kappa_1 \kappa_2 }{(1 + R_2^4 \kappa_2^2 r_2^2)} \dd \varphi_2 \dd r_2 \dd \psi \dd \phi_1 \dd x \nn 
&+& \frac{2 L^4 R_1^2 R_2^4 r_2 \rho r_1^2 \kappa_0 \kappa_1 \kappa_2}{(1- L^4 \kappa_0^2 \rho^2)(1+R_2^4 \kappa_2^2 r_2^2 ) }  \dd t \dd \rho \dd \varphi_2 \dd r_2 \dd \phi_1 - \frac{2 L^3 R_1^3 R_2^3 \rho^2 r_2^2 r_1 \kappa_0 \kappa_1 \kappa_2 }{(1 + R_1^4 \kappa_1^2 r_1^2)} \dd \varphi_1 \dd r_1 \dd \psi \dd \phi_2 \dd x \nn 
&-& \frac{2 L^2 R_1^4 R_2^4 r_2 \rho^2 r_1 \kappa_0 \kappa_1 \kappa_2}{(1+ R_1^4 \kappa_1^2 r_1^2)(1+R_2^4 \kappa_2^2 r_2^2 ) }  \dd \varphi_1 \dd  r_1 \dd \varphi_2 \dd r_2 \dd \psi + \frac{2 L^3 R_1^3 R_2^3 \rho r_2^2 r_1^2 \kappa_0 \kappa_1 \kappa_2 }{(1 - L^4 \kappa_0^2 \rho^2)} \dd t \dd \rho \dd \phi_1 \dd \phi_2 \dd x,  \nonumber
\eea
where once again we have omitted wedge products. 

At this stage, there are two consistency checks that confirm we are on the right track. First, it can be checked that the final $F_5$ is self-dual, a feature we have not put in by hand, rather it falls out of our prescription. Secondly, it can be checked that the Einstein equation $E_{xx} = 0$ is satisfied, once (\ref{radii_constraint}) holds, consistent with our expectation of a solution existing.

To complete the solution, we should use the remaining equations of motion to fix $\phi_0$. With $H=0$, this is most easily done by checking the $B$-field equation (\ref{general2}), which is satisfied once we introduce an additional constant $\kappa$, so that the dilaton shift takes the form, 
\be
e^{2 \phi_0} \equiv 1 + \kappa^2 =  1+ L^4 \kappa_0^2 = 1 + R_1^4 \kappa_1^2 = 1 + R_2^4 \kappa_2^2. 
\ee
This allows us to rewrite the original constants in terms of $\kappa$, 
\be
\kappa_0 = L^{-2} \kappa, \quad \kappa_1 = R_1^{-2} \kappa, \quad \kappa_2 = R_2^{-2} \kappa. 
\ee
With this choice, it is now clear the the dilaton equation (\ref{dilaton_EOM}) is satisfied once (\ref{radii_constraint}) holds. This confirms that a solution exists and it is a lengthy, but straightforward calculation, to check that the Einstein equation also holds. As stated earlier, it is straightforward to generalise the above deformation to a two-parameter deformation. 

\section{Acknowledgements}
We thank Ilya Bakhmatov and Shahin Sheikh-Jabbari for feedback on results, as well as Ben Hoare, Kostas Sfetsos, Dan Thompson, Jun-ichi Sakamoto, Stijn van Tongeren and Kentaroh Yoshida for discussion on related topics. E. \'O C thanks the Theoretical Physics Department at Maynooth University and the Universidad de Murcia for hospitality during the workshop ``Geometry, Duality and Strings 2018", 23-26 May 2018, Murcia, Spain. The work of HY is supported in part by National Natural Science Foundation of China, Project 11675244.

\appendix 

\section{Generalised Supergravity}
The equations of motion of generalised IIB supergravity may be expressed as \cite{Arutyunov:2015mqj,Wulff:2016tju}: 
\bea
R_{MN}-\frac{1}{4}H_{MKL}H_N{}^{KL} - T^{IIB}_{MN} + \nabla_M X_N+\nabla_N X_M = 0, 
\label{general1} \\
\frac{1}{2} \nabla^K H_{KMN}+\frac{1}{2}\mathcal{F}^K\mathcal{F}_{KMN}+\frac{1}{12}\mathcal{F}_{MNKLP}\mathcal{F}^{KLP} = X^K H_{KMN}+\nabla_M X_N-\nabla_N X_M\,, \label{general2}\\
R-\frac{1}{12}H^2+4 \nabla_MX^M-4X_MX^M=0\,,  
\label{general3}
\eea
where we have defined the one-form in terms of the dilaton $\phi$, Killing vector $I$ and $B$-field, 
\be
\label{X}
X = \dd \phi + I + i_{I} B,  
\ee
and the stress-energy tensor, 
\bea
T^{IIB}_{MN} &\equiv &\frac{1}{2}\mathcal{F}_M\mathcal{F}_N
+\frac{1}{4}\mathcal{F}_{MKL}\mathcal{F}_N{}^{KL}
+\frac{1}{96}\mathcal{F}_{MPQRS}\mathcal{F}_N{}^{PQRS} \nn
&-& \frac{1}{4}G_{MN} (\mathcal{F}_K\mathcal{F}^{K}
+\frac{1}{6}\mathcal{F}_{PQR}\mathcal{F}^{PQR})\,, 
\eea
where the forms $\mathcal{F}$ are the usual RR field strengths $F$ rescaled by the dilaton, $\mathcal{F} = e^{\phi} F$. It should be noted that the modification from usual IIB supergravity is captured by the Killing vector and setting $I =0$, we recover the original theory. 

To simplify the equations involving the RR sector only, following \cite{Araujo:2017enj} we introduce Page forms, 
\bea
\label{Page}
Q_1 &=& F_1, \quad Q_3 = F_3 + B \wedge F_1, \quad Q_5 = F_5 + B \wedge F_3 + \frac{1}{2} B^2 \wedge F_1, \nn
Q_7 &=& - * F_3 + B \wedge F_5 + \frac{1}{2} B^2 \wedge F_3 + \frac{1}{3!} B^3 \wedge F_1, \nn
Q_9 &=& * F_1 - B \wedge * F_3 + \frac{1}{2} B^2 \wedge F_5 + \frac{1}{3!} B^3 \wedge F_3 + \frac{1}{4!} B^4 \wedge F_1, 
\eea
so that the equations of motion take the simple form,  
\be
\label{dQQ}
\dd Q_{2 n-1} = i_{I} Q_{2 n +1}, \quad n = 1, 2, 3, 4. 
\ee 
 It is worth noting that the RR sector acts largely as a spectator and the modification is more pronounced in the NS sector, where it may be traced to non-Abelian T-duality with respect to non-semisimple groups \cite{Elitzur:1994ri, Hong:2018tlp}. 
 
The equations of motion of generalised IIA supergravity may be expressed as 
\bea
R_{MN}-\frac{1}{4}H_{MKL}H_N{}^{KL} - T^{IIA}_{MN} + \nabla_M X_N+\nabla_N X_M = 0, \\
\frac{1}{2} \nabla^K H_{KMN}+\frac{1}{2}\mathcal{F} \mathcal{F}_{MN}+\frac{1}{4}\mathcal{F}^{PQ}\mathcal{F}_{MN PQ} = X^K H_{KMN}+\nabla_M X_N-\nabla_N X_M\,,
\eea
with no change for the dilaton equation (\ref{general3}) and we have defined the stress-energy tensor: 
\bea
T^{IIA}_{MN} &\equiv &\frac{1}{2}\mathcal{F}_{M P}\mathcal{F}_N^{~P}
+\frac{1}{12}\mathcal{F}_{MPQR}\mathcal{F}_N{}^{PQR}  \nn
&-& \frac{1}{4}G_{MN} (\mathcal{F}^2 + \frac{1}{2} \mathcal{F}_{PQ} \mathcal{F}^{PQ}+\frac{1}{24}\mathcal{F}_{PQRS}\mathcal{F}^{PQRS}), \
\eea
where the zero, two and four-form are related to the usual RR field strengths through a factor of the dilaton, $\mathcal{F} = e^{\phi} F$. Defining the Page forms, 
\bea
Q_0 &=& F_0, \quad Q_2 = F_2 + B F_0, \quad Q_4 = F_4 + B \wedge F_2 + \frac{1}{2} B^2 F_0, \nn
Q_6 &=& - * F_4 + B \wedge F_4 + \frac{1}{2} B^2 \wedge F_2 + \frac{1}{3!} B^3 F_0,  \nn
Q_8 &=&  * F_2 - B \wedge * F_4 + \frac{1}{2} B^2 \wedge F_4 + \frac{1}{3!} B^3 \wedge F_2 + \frac{1}{4!} B^4 \wedge F_0. 
\eea
Note that in contrast to usual massive IIA supergravity, $F_0$ may not be a constant. The equations of motion of the RR sector take the form 
\be
\dd Q_{2 n} = i_{I} Q_{2 n +2}, \quad n = 0, 1, 2, 3, 4. 
\ee

\section{A trivial solution}
In \cite{Wulff:2018aku} a class of ``trivial" solutions to generalised supergravity were identified. Solutions are deemed trivial when they satisfy the standard supergravity equations, even though $I \neq 0$,  and this happens when the following two conditions are met:
\be
\label{trivial_condition}
I_{\mu} I^{\mu} = 0, \quad \dd I = i_{I} H. 
\ee 
Here we present a simple example of a trivial solution by considering a deformation of flat spacetime. In particular, we consider flat spacetime in 3D \footnote{We thank S. van Tongeren for pointing out the 2D deformation is simply flat spacetime, thus making it completely trivial.}, but this example immediately generalises to higher dimensions. We recall the $r$-matrix (\ref{r-matrix}) and take the vector $a$ to be null, $a  = (\eta, \eta, 0)$. See \cite{Hoare:2016hwh} for an earlier pp-wave example of a trivial solution. 

Concretely, we consider the spacetime, 
 \be
 \dd s^2 = - \dd t^2 + \dd x^2 + \dd y^2,   
 \ee
 and we will use our prescription outlined in section \ref{sec:method} to generate the deformed geometry. Replacing the generators by Killing vectors, while identifying $\Theta = r$, the deforming bivector becomes: 
 \be
 \Theta^{tx}  = \eta ( t+x ), \qquad  \Theta^{ty}  = \eta y,\qquad  \Theta^{xy}  = -\eta y. 
 \ee
 
Using (\ref{open_closed}), (\ref{I_theta}) and (\ref{tduality_inv}), the deformed solution is  
\bea
\dd s^2 &=&  \frac{1}{\Delta} \bigg[ - (1+\eta^2y^2)\dd t^2 + (1-\eta^2y^2) \dd x^2 + \dd y^2 -2\eta^2y^2 \dd t\dd x \nn
&& + 2\eta^2(t+x)y\dd t \dd y +2\eta^2(t+x)y \dd x \dd y  \bigg],  \nn
B &=& \frac{1}{\Delta} \left[\eta (t+x) \dd t \wedge \dd x + \eta y \dd t \wedge \dd y + \eta y \dd x \wedge \dd y  \right], \nn
\Phi &=& - \frac{1}{2} \log \Delta, \quad I = 2 \eta ( -\partial_t + \partial_x), 
\eea
where we have defined 
\be
\Delta = 1 - \eta^2 ( t + x)^2. 
\ee
Lowering the indices on the Killing vector, the one-form $I$ is 
\be
I = \frac{2 \eta}{\Delta} ( \dd t + \dd x). 
\ee
It is easy to check that $I$ is null and since both it and $B$ are closed, the second condition presented in (\ref{trivial_condition}) is trivially satisfied. As a result, this solution constitutes an example of a trivial solution. This means that the dilaton can be shifted to absorb the Killing vector $I$, so that the deformed spacetime may be regarded as a solution to standard supergravity.

\end{document}